\documentstyle[12pt]{article}
\newcommand{\ba}{\begin{eqnarray}}
\newcommand{\ea}{\end{eqnarray}}
\newcommand{\be}{\begin{equation}}
\newcommand{\ee}{\end{equation}}
\newcommand{\ed}{\end{document}}
\newcommand{\lab}[1]{\label{#1}}
\newcommand{\re}[1]{(\ref{#1})}
\newcommand{\ci}[1]{\cite{#1}}
\renewcommand{\baselinestretch}{1.3}
\begin{document}\large
\date{}
\title{ SPECTRA OF DOUBLY HEAVY QUARK BARYONS}
\author{D. U. MATRASULOV,\\
Heat Physics Department of the Uzbek Academy of Sciences,\\
28 Katartal St., 700135 Tashkent, Uzbekistan\\
M. M. MUSAKHANOV\\
Theoretical Physics Department, Uzbekistan National University\\
Vuzgorodok, 700174 Tashkent, Uzbekistan\\
e-mail: yousuf@iaph.silk.org
\\
and\\
T. MORII\\
Faculty of Human Development, Kobe University\\
3-11 Tsurukabuto, Nada, Kobe 657-8501, Japan}

\maketitle
\begin{abstract}
Baryons containing two heavy quarks are treated in the Born-Oppenheimer
approximation. Schr\"odinger equation for two center Coulomb plus
harmonic oscillator potential is solved by the method of ethalon equation at
large intercenter separations. Asymptotical expansions for energy term and
wave function are obtained in the analytical form. Using those formulas,
the energy spectra of doubly heavy baryons with various
quark compositions are calculated analytically.

\end{abstract}
PACS numbers: 03.65.Pm, 12.39.Pn, 12.40 Yx.\\

\section*{Introduction}

The investigation of properties of hadrons containing one or more heavy
quarks is very important for understanding the dynamics of quark and gluon
interactions. Presently at the LHC, B-factories and the Tevatron with high
luminosity, several experiments have been proposed, in which a detailed
study of baryons containing two heavy quarks can be performed. In particular,
in the forthcoming experiment at CERN, the COMPASS group is going to
find the doubly charmed baryons and study their physical properties. In this
connection, doubly heavy quark baryons are now becoming one of the most
exciting subjects in particle physics.  Therefore, theoretical predictions of
properties of doubly heavy quark baryons acquire a big significance for
the forthcoming experimental study of these particles. So far there
have been various approaches by which their mass spectra and other
properties can be calculated. One of them is the nonrelativistic
quark model which gives
relatively accurate results for baryon spectra \ci{Rich89,Rich92,Mukh93}.
The possible quark compositions of doubly heavy quark baryons are
$ccq$, $cbq$ and $bbq$, where $q$ denotes a light $u$, $d$ or $s$ quark.
Note that the baryons containing the top quark(s) are not practical
subject here because a top quark is extremely heavy and hence we have no
chance to find them as stable hadrons.  The doubly heavy quark
baryons may be considered as an
analogue of the hydrogen molecular ion $H^+_2$, which has been treated
successfully in the Born-Oppenheimer approximation.  The same approximation
is expected to be efficient even for doubly heavy quark baryons,
though there exist some differences between these baryons and $H^+_2$
systems. One of them is, for this case, the
appearance of the confining potential in addition to the QCD Coulomb
potential.  As is well known, the variables of the Schr\"odinger equation with
two-center Coulomb plus confining potential cannot be separated
for their kinematical variables, in general.
To our knowledge, the two-center potential which allows the
separation of variables is only the two-center Coulomb
plus harmonic oscillator potential.

In this paper, we treat $QQq$ baryons in
the nonrelativistic approach by using the solution of the
Schr\"odinger equation with two-center
Coulomb plus harmonic oscillator potential, i. e.
the well-known method of ethalon equation which is widely used for solving
Schr\"odinger equation with two-center pure Coulomb potential in the physics
of $H^{+}_{2}$ \ci{Kom78,Slav,Kom68}. First, we give a general
scheme of treatment $QQq$ baryon in the Born-Oppenheimer approximation.
Then, the two-center Schr\"odinger equation
with two-center Coulomb plus harmonic oscillator potential is
analytically solved with some approximation: the energy term of
the light quark moving in the field of two
heavy quarks is obtained in the form of asymptotical
expansion over the inverse power
of the distance between heavy quarks. Finally, we give an analytical
formula of the baryon energy spectrum for $QQq$.

\section*{Doubly heavy quark baryon in the Born-Oppenheimer approximation}
In the Born-Oppenheimer approximation the wave function
is split into heavy- and light-quark degrees of freedom
$$
\Psi(R,r) = \sum_{n}\phi_{n}(R)\psi_{n}(R,r),
$$
where $R$ is the distance between two heavy quarks and $r$
is the distance between light quark and center-of-mass
of the heavy-quark pair. The light quark wave function $\psi(r,R)$ and
its energy term $E(R)$ can be found from the Schr\"odinger equation
$$
[-\frac{1}{2m_{q}}\Delta + V(r_{1}) +V(r_{2})]\psi=E(R)\psi,
$$
where $r_{1}$ and $r_{2}$ are the distances between light and heavy
quarks, $Q_1$ and $Q_2$, respectively.
The binding energy of this system is approximated by the equation:
$$
[-\frac{1}{2\bar M_{QQ}}\Delta + V_{QQ}(R) + E(R)]\phi=\varepsilon\phi,
$$
where $\bar M_{QQ}$ is the reduced mass of $QQ$.

A quark potential with Coulomb plus harmonic confinement for this baryon
is given by\ci{Mukh93}
$$
V(r_{ij}) = \sum\nolimits_{i,j} \frac{1}{4}\lambda_{i}\lambda_{j}(V_{0} - Ar_{ij}^2 +
\frac{\alpha_{s}}{r_{ij}}) =
-\frac{2}{3}\sum\nolimits_{i,j}(V_{0} - Ar_{ij}^2 +
\frac{\alpha_{s}}{r_{ij}})
$$
In the field of two heavy quarks with this potential, the motion of
a light quark can be nonrelativistically described by the following
Schr\"odinger equation,
\be
[-\frac{1}{2}\Delta -
\frac{Z}{r_{1}}-\frac{Z}{r_{2}}+\omega^{2}(r_{1}^2+r_{2}^2)-
\frac{4}{3}V_{0}]\psi=E(R)\psi,
\lab{eq1}
\ee
where $Z=2\alpha_{s}/3$ and $\omega^2 = 2A/3$.

In the prolate spheroidal coordinates defined as
$$\xi=\frac{r_{1}+r_{2}}{R}\,\, (1<\xi<\infty),\,\,\,
 \eta=\frac{r_{1}-r_{2}}{R}\,\, (-1<\eta<1),$$
the potential term in eq.\re{eq1} can be written in the form
\be V(r_{1},r_{2})= -\frac{2}{R^2}
\frac{a(\xi)+b(\eta)}{\xi^2-\eta^2}+\frac{\omega^2R^2}{2} -\frac{4}{3}V_{0},
\lab{pot}
\ee
where
$$a(\xi)=2ZR-\frac{\omega^2R^4}{4}\xi^2(\xi^2-1),\,\,
b(\eta)=2ZR-\frac{\omega^2R^4}{4}\eta^2(\eta^2-1).$$

As is well known\ci{Kom78}, the Schr\"odinger equation with the potential in
the form of eq.\re{pot} is separable in the prolate spheroidal coordinates.
Then it is convenient to use
$$
\psi = \frac{U(\xi)}{\sqrt{\xi^2-1}}\frac{V(\eta)}{\sqrt{1-\eta^2}}
\frac{e^{\pm im\phi}}{\sqrt{2\pi}},
$$
where $\phi$ and $m$ are azumuthal angle and azimuthal quantum number, respectively.
After substituting this into eq.\re{eq1},
we obtain from the following ordinary
differential equations connected
with separation constants $\lambda$ and $m$:

\be
U''(\xi)+[\frac{h^2}{4}+\frac{h(\alpha\xi-\lambda)}{\xi^2-1}-h^4\gamma\xi^2
+\frac{1-m^2}{(\xi^2-1)^2}]U(\xi)=0,
\lab{rad}
\ee
\be
V''(\eta)+[\frac{h^2}{4}+\frac{h\lambda}{1-\eta^2}-h^4\gamma\eta^2
+\frac{1-m^2}{(1-\eta^2)^2}]V(\eta)=0,
\lab{ang}
\ee
where $\alpha=2Z/\sqrt{2E'}$ and $\gamma=\omega^2/8E'^2$, and further
\be
h=\sqrt{2E'}R,
\lab{energy}
\ee
with
$$
E' = E -\frac{\omega^2R^2}{2}+\frac{4}{3}V_{0}.
$$
Finiteness and continuity of the wave function $\psi$
in the whole space lead to the following boundary conditions
for the functions $U$ and $V$:
\be U(\xi)\mid_{\xi=1} =0 ,\;\;\; U(\xi)\mid_{\xi \longrightarrow \infty}
\longrightarrow 0 ,
\lab{radbound}
\ee
\be V(\eta)\mid_{\eta =\pm 1} = 0.
\lab{angbound}
\ee

\section*{Asymptotics of quasi-angular equation}
We will approximately solve eqs.\re{rad} and\re{ang} for
large $R$ by the
method of ethalon equation. This method is successfully applied to the
solution of nonrelativistic two center Coulomb problem \ci{Kom78,Slav,Kom68}
and in the theory of diffraction of waves.  Details on the method of
ethalon equation are given in \ci{Kom78,Slav,Kom68,Kom67}, also briefly described in Appendix.

Let us start from the angular equation of \re{ang}. As an ethalon
equation for eq.\re{ang}, we choose the following
Whittaker equation \ci{Erd}:  \be
 W''+[-\frac{h^4}{4}+\frac{h^2k}{z}+\frac{1-m^2}{4z^2}]W \lab{ethalon1}=0
\lab{whit}
 \ee
and seek a solution  in the form
\be V=[z'(\eta)]^{-\frac{1}{2}}M_{k,\frac{m}{2}}(h^2z),
\lab{sol1}
\ee
where $M_{k,\frac{m}{2}}(h^2z)$ is the solution (regular at zero) of
eq.\re{ethalon1}. Substituting \re{sol1} into \re{ang} and taking into
account \re{whit}, we get the following equation for $z$:
\ba
\frac{z'^2}{4}-\gamma(x-1)^2 - \frac{1}{h^2}(\frac{1}{4}
+\frac{kz'^2}{z} -\frac{\lambda}{2x(1-x/2)})+
\nonumber \\
\frac{\tau}{h^2}(\frac{1}{x^2(1-x^2)} -\frac{z'^2}{z^2})
-\frac{1}{2h^2}\{z,x\}=0,
\lab{zeq}
\ea
where
$$\{z,x\}= -\frac{3}{2}(\frac{z''}{z'})^2+\frac{z'''}{z'},$$
and {$\tau=\frac{1-m^2}{4},$ $x=1+\eta$.

Requirement of coincidence
at the transition points \ci{Kom68,Kom67}
$$ z(x)\mid_{x=0}=0, $$
leads to the following "quantum condition"
\be \lambda = 2kz'(0)+ \frac{2\tau}{h^2}[\frac{z''(0)}{z'(0)} -1].
\lab{lambda}
\ee
We will seek the solution of eq.\re{zeq} and eigenvalues $\lambda$ in the
form of following asymptotical expansion:  $$ z=\sum^{\infty}_{k=0}
\frac{z_{k}}{h^k},\;\;\;\;\;
\lambda=\sum^{\infty}_{k=0}\frac{\lambda_{k}}{h^k}. $$

Substitution of these expansions into \re{zeq} gives us the recurrence
system of differentil equations for $z$ :
$$z'_{0}=2\gamma^{\frac{1}{2}}(x-1), $$
$$z'_{1}=0, $$
$$ z'_{2}=\frac{1}{2z'_{0}}+\frac{2kz'_{0}}{z_{0}} -
\frac{(z'_{1})^2}{2z'_0} - \frac{2\lambda_0}{z'_0x(1-x/2)} -
\frac{z'^2_2}{2}, $$
$$ .\; .\; .\; .\; .\; .\; .\; .\; .\; .\; .\; .\; .\; .\; .\; .\; .\; .,$$
and for $\lambda$ :  $$\lambda_0 =2kz'_0(0),$$
$$\lambda_1 = 2kz'_1(0), $$ $$\lambda_2 = 2kz'_2(0) +
2\tau(\frac{z''_0(0)}{z'_1(0)} - 1),$$ $$ .\;\; .\; .\; .\; .\; .\; .\; .\;
.\; .\; .\; .\; .\; .\; ..$$
Solving these recurrence equations, we obtain
\be \lambda^{(\eta)} = 4k\gamma^{\frac{1}{2}} + \frac{2k\beta - 4\tau}{h^2}
+ O(\frac{1}{h^4}),
\lab{lambda1}
\ee
for $\lambda$, and
\be z= \gamma^{\frac{1}{2}}x(2-x) + \frac{1}{h^2}\beta ln(1-x)
+ O(\frac{1}{h^4}),
\lab{wave1}
\ee
for $z$.

>From boundary conditions one can obtain for quantum number
$k$\ci{Slav,Kom68}:
$$ k = q + \frac{m+1}{2}, $$  where $q = 0,1,2,.... $.

\section*{Asymptotics of quasi-radial equation}
As an ethalon equation for eq.\re{rad}, we take the following
equation :
\be W''+[h^2s - h^4y^2 - \frac{4\tau+3}{4y^2}]W =0, \ee
a solution of which is expressed by the confluent hypergeometric
functions\ci{Erd,Abr}
$$ W = y^c e^{-\frac{h^4y^2}{2}}
F(\frac{s-2c -1}{4}, c+\frac{1}{2}, h^4y^2),$$
where $c = \frac{1+\sqrt{m^2+3}}{2}$.

Boundary condition \re {radbound} and properties of functions $F$
\ci{Erd} give rise to the  following expression for $s$:
 $$ s = 4n + \sqrt{m^2 +3} +2 .$$
Substituting $$ U = [y(\xi)]^{-\frac{1}{2}}W(y(\xi)) $$ into
eq.\re{rad}, we obtain
\be \frac{y^2y'^2}{4} - \gamma\xi^2 +
\frac{1}{ h^2}(\frac{1}{4} - sy'^2 - \frac{\lambda}{\xi^2 -1})
+\frac{1}{h^3} \frac{\alpha\xi}{\xi^2 -1} + \frac{4\tau}{h^4(\xi^2-1)^2} -
\frac{3-4\tau}{4h^4}\frac{y'^2}{y^2} -\frac{1}{2h^4} \{y,\xi\} = 0.
\lab{nonr}
\ee

After substitution of $$\phi = \frac{y^2(t)}{4},$$ this equation
can be reduced to the form
\be
\phi'^2 -\gamma(t+1) +\frac{1}{h^2}(\frac{1}{4}-(n+\frac{1}{2})
\frac{\phi'^2}{\phi}-\frac{\lambda}{t(t+2)})+\frac{1}{h^3}\;\frac{\alpha
(t+1)}{t(t+2)} \nonumber \\ +\frac{\tau}{h^4}(\frac{\phi'^2}{\phi^2} -
\frac{4}{t^2(t+2)^2}) - [\phi,t]= 0,
\lab{nonlin}
\ee
where $t = \xi -1.$
The quantization condition which follows from $\phi(x) =0$ is
written in the form
\be
\lambda = -2s\phi'(0) + \frac{\alpha}{h} -\frac{1}{h^2}
[\frac{\phi''}{\phi'} + 1]\mid_{t=0}.
\ee
Inserting the asymptotical expansions
$$ \phi=\sum^{\infty}_{k=0} \frac{\phi_{k}}{h^k},\;\;\;\;\;
\lambda=\sum^{\infty}_{k=0}\frac{\lambda_{k}}{h^k} $$
into eq.\re{nonlin} and solving the equations obtained herewith,
we get the following result
\ba
y = 2\gamma^{\frac{1}{4}}(t^2+2t)^{\frac{1}{2}} + \frac{1}{h^2}
\delta\gamma^{-\frac{1}{4}}(t^2+2t)^{-\frac{1}{2}}ln(t+1) +
\nonumber \\
\frac{1}{h^3}\alpha\gamma^{-\frac{3}{4}}(t^2+2t)^{-\frac{1}{2}}
ln\frac{2(t+1)}{t+1}+ O(\frac{1}{h^4}),
\lab{wave2}
\ea
for $y$, and
\be \lambda^{(\xi)} = -2s\gamma^{\frac{1}{2}} - \frac{\alpha}{h} +
\frac{4\tau -s\delta}{h^2} - \frac{s\alpha \gamma^{-\frac{1}{4}}}{2h^3} +
 O(\frac{1}{h^4}),
\lab{lambda2}
\ee
for $\lambda$.

\section*{Asymptotical expansion for energy}
Asymptotical expansions \re{lambda1} and \re{lambda2} give us an expression
for the energy term in the form of multipole expansion.
In order to obtain this expansion one should insert
$$ E' = E_{0} + \frac{E_{1}}{R} + \frac{E_{2}}{R^2} + ...  $$
into eqs.\re{lambda1} and \re{lambda2}.
Equating $\lambda^{(\eta)}$ to be $\lambda^{(\xi)}$ and taking into account
\re{energy}, we get the following equations for coeficients $E_{1}, E_{2},
 ... $ :

$$ E_{1} = \frac{1}{6Z}[(s\omega-2k\omega^{-1})(2E_{0})^{\frac{5}{2}} +
 (4s^2 -16k^2 -16\tau)(2E_{0})^{\frac{3}{2}}] , $$
$$ E_{2} = \frac{5}{2}E_{1}^2 + 2s\omega^{-1}E_{0} +
E_{1}(2E_{0})^{\frac{1}{2}}Z^{-1}(16\tau^2 + 16k^2 - 4s^2), $$
$$ .\,.\,.\,.\,.\,.\,.\,.\,.\,.\,.\,.\,.\,.$$

Now we need to find $E_{0}$. In order to find this value, we note that
for $ R \longrightarrow \infty, $  $E' = E_{0}$ and hence we have
\be E = E_{0} + \frac{\omega^2R^2}{2}+\frac{4}{3}V_{0}.
\lab {energy2}
\ee
On the other hand, for large $R$ we have
\ba
V(r_{1}, r_{2}) = \frac{2Z}{R}\sum^{\infty}_{l=0}(\frac{r}{R})^l
 P_{l}(cos\theta) + \omega^2[(r^2 +2rRcos\theta +\frac{R^2}{4}) +
\nonumber \\
(r^2 - 2rRcos\theta +\frac{R^2}{4})]\approx
 \omega^2(2r^2 + \frac{R^2}{2})-\frac{4}{3}V_{0}.
\ea
Hence, for the energy term with this potential  we obtain
\be E = 2\omega(N + \frac{3}{2}) + \frac{\omega^2R^2}{2}+\frac{4}{3}V_{0},
\lab{energy3}
\ee
 where $N = n +q + m+1 $ is the principial quantum number. Comparing
eqs.\re{energy2} and \re{energy3}, we obtain
$$ E_{0} = 2\omega(N +\frac{3}{2}).$$

Thus, the following asymptotical expansion is obtained for the energy
term of light quark in the field of two heavy quarks:
$$
 E =-\frac{4}{3}V_{0}+\frac{\omega^2R^2}{2}+E_{0} + \frac{E_{1}}{R} + \frac{E_{2}}{R^2} + ...
$$
%\newpage
\section*{$QQq$ baryon spectra}

As mentioned above, the $QQq$ binding energy can be finally
obtained by solving the Schr\"odinger equation
\be
[-\frac{1}{2\bar M_{QQ}}\Delta + V_{QQ}(R) + E(R)]\phi=\varepsilon\phi.
\lab{equ2}
\ee
If one takes $E(R)$ in the form
$$
 E =-\frac{4}{3}V_{0}+\frac{\omega^2R^2}{2}+E_{0} + \frac{E_{1}}{R},
$$
for
$$
V_{QQ}(R) = \omega^2 R^2 - \frac{Z}{R}- \frac{2}{3}V_{0},
$$
then eq.\re{equ2} can be rewritten as
\be
[-\frac{1}{2\bar M_{QQ}}\Delta + \omega'^2 R^2 - \frac{Z'}{R}-V'_{0}]\phi=\varepsilon\phi,
\lab{equ3}
\ee
where $Z'= Z-E_{1},$  $\omega'^2 = \frac{3}{2}\omega^2$
and $V_{0}' =2V_{0}-E_0$.

To solve this equation, we use the result of \ci{Fran} where a method
for an analytical solution of the Schr\"odinger
equation with potential
$$
V(R) = -\frac{Z}{R} +\lambda R^k
$$
was offered. Details on this method and its application to our potential
are given in Appendix.
Application of this method to eq. \re{equ3} gives us
\be
\varepsilon_{Nnl} = 2\omega(N+\frac{3}{2}) +[Z'^2\omega'^6r_{nl}]^{1/5}-2V_0,
\lab{spectrum}
\ee
where $N$ is the principial quantum number of the light
quark moving in the field
of $QQ$, $r_{nl}$ is defined in Appendix.
The formula \re{spectrum} describes
the energy spectrum of the $QQq$ baryon. In tables 1, 2 and 3,
the mass spectra of $ccq$, $bbq$ and $bcq$ baryons calculated
using the formula \re{spectrum} are given, respectively. The following values of
potential parameters are choosen in this calculation $\alpha_s = 0.39, \;$ $\omega^2 = 0.174 GeV^3,\;$
$V_0 = 0.05 GeV$ for the potential
$$
V = \frac{2}{3}(-\frac{\alpha_s}{r} +\omega^2 r - V_0).
$$

\section*{Conclusion}
In this work we have treated doubly heavy baryons in the
Born-Oppenheimer approximation. The following two problems have been solved
in the framework of this approximation: (1)the Schr\"odinger equation
for two-center Coulomb plus harmonic oscillator potential and (2)the
Schr\"odinger equation for central symmetric Coulomb plus harmonic
oscillator potential. As the final result an analytical
formula for the energy spectrum
of baryons containing two heavy quarks is derived.  Obtained
formula is applied for the calculation of mass spectra of doubly
heavy quark baryons with various quark compositions.
The above analytical results could be useful for further
numerical calculations in
non-asymptotical region.

\section*{Appendix}
\section{The scaling variational method and its application to Coulomb
plus confining potential}

Consider the following Hamiltonian
\be
H = -\frac{1}{2}\Delta +V(r),
\ee
which obeys the eigenvalue equation
\be
H\psi_{nl} = E_{nl}\psi_{nl}, \;\; <\psi_{nl}\mid\psi_{n'l'}> =\delta_{nn'} \delta_{ll'},
\ee
$$n,n' = 1,2,..,\;\;\;l,l' = 1,2,.., $$
where $n$ and $l$ denote the principal and angular quantum numbers, respectively.
To solve this equation, we start from a set of functions $\{\phi_{nl}\}$
which are the eigenfunctions of an arbitrary central field Hamiltonian $H_0$:
\be
H_0\phi_{nl} =\epsilon_{nl}\phi_{nl}, \;\;\;<\phi_{nl}\mid\phi_{n'l'}> =\delta_{nn'} \delta_{ll'}.
\ee
Then, we construct the functionals
\be
\varepsilon_{nl}(\alpha) = <\phi_{nl}^{\alpha}\mid H\phi_{nl}^{\alpha}>,
\ee
where
$$
\phi_{nl}^{\alpha} = \alpha^{3/2}\phi_{nl}(\alpha r).
$$
The value for the $\alpha$ is determined by
\be
(\frac{\partial \varepsilon}{\partial\alpha})(\alpha =a) = 0.
\ee

Let us now apply this method to our problem, i.e. to
the Schr\"odinger equation with potential
$$
V(r) =  - \frac{Z'}{R}+\omega'^2 r^2.
$$
As $H_0$, we choose pure Coulomb Hamiltonian, i.e.
$$
H_0 = -\frac{1}{2}\Delta - \frac{Z'}{R}.
$$
Then, $\epsilon_{nl} = -\frac{Z'^2}{2n^2}$ with $n =n_r+l+1$, where
$n_r$ is the radial quantum number.

According to the above procedure, we have
\begin{eqnarray}
\displaystyle
\varepsilon_{nl}(\alpha) = <\phi_{nl}^{\alpha}\mid H_0\mid\phi_{nl}^{\alpha}>
+<\phi_{nl}^{\alpha}\mid\omega^2r^2\mid \phi_{nl}^{\alpha}>=\\ \nonumber
 -\frac{Z'^2\alpha^3}{2n^2} + \frac{\omega^2\alpha^{-2}n^2}{2}[5n^2+1-3l(l+1)],
\end{eqnarray}
For calculation of the second matrix element, we have used the well-known
expression for average value $\bar r^2$ in Coulomb field
which is given in \ci{Land}.
>From $\partial \varepsilon (\alpha)/\partial \alpha = 0$, we obtain
$$
\alpha_0 = \{-\frac{2\omega^2n^4}{3Z^2}[5n^2+1-3l(l+1)]\}^{1/5}.
$$
Then, for the energy level one can get the following analytical formula
\be
\varepsilon_{nl} = [Z^2\omega^6r_{nl}]^{1/5},
\ee
where
$$
r_{nl} = n^2[5n^2+1-3l(l+1)]^3.
$$
\newpage
\section{The formal procedure of the method of ethalon equation}
Let's consider the following second order differential equation:
\be
y''(x) + p^2[\lambda-q(x)]y(x) = 0
\lab{dif}
\ee
in the interval $[a,b]$. Let in this interval eq.\re{dif} has one transition
point(poles and zeros of function $Q(x,\lambda) = q(x)-\lambda$ are called
transition points of this equation).

Equation
\be
w''(z) -p^2R(z)w(z) = 0
\lab{ethalon}
\ee
which has the same or close transition points as eq.\re{ethalon}
is called the ethalon equation for eq.\re{dif}.

Solution of eq.\re{dif} we will seek in the form
\be
y(x) = [z'(x,p)]^{-1/2}w(z(x,p))
\lab{sol}
\ee
where $w$ is the solution of eq.\re{ethalon}.
Inserting \re{sol} in to eq.\re{dif} and taking into account eq.\re{ethalon}
we obtain the following (nonlinear) differential equation for $z(x,p)$:
\be
R(z)z'^2 - Q(x,\lambda) - \frac{1}{2p^2}\{z,x\} = 0
\ee
 here
$$\{z,x\}= -\frac{3}{2}(\frac{z''}{z'})^2+\frac{z'''}{z'},$$
In the case of eq. \re{ang} we have for $Q$
$$
Q = -[\frac{h^2}{4}+\frac{h\lambda}{1-\eta^2}-h^4\gamma\eta^2
+\frac{1-m^2}{(1-\eta^2)^2}]
$$
and for $R$ (from eq.\re{whit}
$$
R =-[-\frac{h^4}{4}+\frac{h^2k}{z}+\frac{1-m^2}{4z^2}].
$$
So, for $z$ one obtains eq.\re{zeq}.
$$
\frac{z'^2}{4}-\gamma(x-1)^2 - \frac{1}{h^2}(\frac{1}{4}
+\frac{kz'^2}{z} -\frac{\lambda}{2x(1-x/2)})+
\nonumber \\
\frac{\tau}{h^2}(\frac{1}{x^2(1-x^2)} -\frac{z'^2}{z^2})
-\frac{1}{2h^2}\{z,x\}=0
$$
\newpage

Table 1. The mass spectrum of $ccq$ baryon (in GeV) calculated using the
formula \re{spectrum}; $m_q =0.385$ GeV, $m_c =1.486$ GeV, $n_l$ and $n_d$
are the principial quantum numbers of light quark and $cc$ diquark,
respectively, $L$ is the orbital quantum number of $cc$ diquark.
\vskip 1cm
\begin{tabular}[3]{c|c|c|c|c|c|c|c|c|c|}
\hline
&$n_{l},n_{d},L$& mass&$n_{l},n_{d},L$ & mass&$n_{l},n_{d},L$ &mass&$n_{l},n_{d},L$ &mass\\  \hline
& 1,1,0 &3.661 & 1,1,1 &3.613  &1,2,2 &3.649 &1,3,3 &3.694 \\
& 1,2,0 &3.730 & 1,2,1 &3.708  &1,3,2 &3.764 &1,4,3 &3.825 \\
& 1,3,0 &3.816 & 1,3,1 &3.799  &1,4,2 &3.872 &1,5,3 &3.949 \\
& 1,4,0 &3.914 & 1,4,1 &3.901  &1,5,2 &3.988 &1,6,3 &4.077 \\
& 1,5,0 &4.024 & 1,5,1 &4.012  &1,6,2 &4.110 &1,7,3 &4.210 \\
& 2,1,0 &3.839 & 2,1,1 &3.791  &2,2,2 &3.828 &2,3,3 &3.873 \\
& 2,2,0 &3.908 & 2,2,1 &3.887  &2,3,2 &3.942 &2,4,3 &4.003 \\
& 2,3,0 &3.994 & 2,3,1 &3.978  &2,4,2 &4.051 &2,5,3 &4.127 \\
& 2,4,0 &4.093 & 2,4,1 &4.079  &2,5,2 &4.166 &2,6,3 &4.255 \\
& 2,5,0 &4.202 & 2,5,1 &4.190  &2,6,2 &4.289 &2,7,3 &4.388 \\
& 3,1,0 &4.018 & 3,1,1 &3.970  &3,2,2 &4.006 &3,3,3 &4.051 \\
& 3,2,0 &4.086 & 3,2,1 &4.065  &3,3,2 &4.120 &3,4,3 &4.182 \\
& 3,3,0 &4.172 & 3,3,1 &4.156  &3,4,2 &4.229 &3,5,3 &4.305 \\
& 3,4,0 &4.271 & 3,4,1 &4.257  &3,5,2 &4.344 &3,6,3 &4.433 \\
& 3,5,0 &4.380 & 3,5,1 &4.369  &3,6,2 &4.467 &3,7,3 &4.567 \\
\hline
\end{tabular}
\newpage
Table 2. The mass spectrum of $bbq$ baryon (in GeV) calculated using
the formula \re{spectrum}; $m_q =0.385$ GeV, $m_b =4.88$ GeV, $n_l$
and $n_d$ principial quantum numbers of light quark and $bb$ diquark,
respectively, $L$ is the orbital quantum number of $bb$ diquark.
\vskip 1cm
\begin{tabular}[3]{c|c|c|c|c|c|c|c|c|c|}
\hline
&$n_{l},n_{d},L$& mass&$n_{l},n_{d},L$ & mass&$n_{l},n_{d},L$ &mass&$n_{l},n_{d},L$ &mass\\  \hline
& 1,1,0 & 9.890 & 1,1,1 & 9.874  &1,2,2 & 9.886 &1,3,3 & 9.900 \\
& 1,2,0 & 9.911 & 1,2,1 & 9.905  &1,3,2 & 9.922 &1,4,3 & 9.942 \\
& 1,3,0 & 9.939 & 1,3,1 & 9.934  &1,4,2 & 9.957 &1,5,3 & 9.981 \\
& 1,4,0 & 9.970 & 1,4,1 & 9.966  &1,5,2 & 9.993 &1,6,3 & 10.022 \\
& 1,5,0 &10.005 & 1,5,1 &10.001  &1,6,2 &10.032 &1,7,3 & 10.064 \\
& 2,1,0 &10.096 & 2,1,1 &10.081  &2,2,2 &10.093 &2,3,3 &10.107 \\
& 2,2,0 &10.118 & 2,2,1 &10.112  &2,3,2 &10.129 &2,4,3 &10.149 \\
& 2,3,0 &10.146 & 2,3,1 &10.141  &2,4,2 &10.164 &2,5,3 &10.188 \\
& 2,4,0 &10.177 & 2,4,1 &10.173  &2,5,2 &10.200 &2,6,3 &10.229 \\
& 2,5,0 &10.212 & 2,5,1 &10.208  &2,6,2 &10.239 &2,7,3 &10.271 \\
& 3,1,0 &10.303 & 3,1,1 &10.288  &3,2,2 &10.300 &3,3,3 &10.314 \\
& 3,2,0 &10.325 & 3,2,1 &10.318  &3,3,2 &10.336 &3,4,3 &10.356 \\
& 3,3,0 &10.353 & 3,3,1 &10.347  &3,4,2 &10.371 &3,5,3 &10.395 \\
& 3,4,0 &10.384 & 3,4,1 &10.380  &3,5,2 &10.407 &3,6,3 &10.436 \\
& 3,5,0 &10.419 & 3,5,1 &10.415  &3,6,2 &10.446 &3,7,3 &10.478 \\
\hline
\end{tabular}
\newpage

Table 3. The mass spectrum of $bcq$ baryon (in GeV) calculated using
the formula \re{spectrum}; $m_q =0.385$ GeV, $m_b =4.88$ GeV,
$m_c =1.486$ GeV,  $n_l$ and $n_d$ principial quantum numbers
of light quark and $bc$ diquark, respectively, $L$ is the orbital
quantum number of $bc$ diquark.
\vskip 1cm
\begin{tabular}[3]{c|c|c|c|c|c|c|c|c|c|}
\hline
&$n_{l},n_{d},L$& mass&$n_{l},n_{d},L$ & mass&$n_{l},n_{d},L$ &mass&$n_{l},n_{d},L$ &mass\\  \hline
& 1,2,0 &7.217 & 1,2,1 & 7.160 &1,3,2 &7.178 &1,4,3 &7.199 \\
& 1,3,0 &7.259 & 1,3,1 & 7.206 &1,4,2 &7.233 &1,5,3 &7.263 \\
& 1,4,0 &7.307 & 1,4,1 & 7.251 &1,5,2 &7.286 &1,6,3 &7.324 \\
& 1,5,0 &7.361 & 1,5,1 & 7.300 &1,6,2 &7.343 &1,7,3 &7.386 \\
& 2,1,0 &7.438 & 2,1,1 & 7.355 &2,2,2 &7.403 &2,3,3 &7.452 \\
& 2,2,0 &7.471 & 2,2,1 & 7.414 &2,3,2 &7.432 &2,4,3 &7.454 \\
& 2,3,0 &7.513 & 2,3,1 & 7.461 &2,4,2 &7.488 &2,5,3 &7.518 \\
& 2,4,0 &7.562 & 2,4,1 & 7.505 &2,5,2 &7.541 &2,6,3 &7.579 \\
& 2,5,0 &7.615 & 2,5,1 & 7.555 &2,6,2 &7.597 &2,7,3 &7.641 \\
& 3,1,0 &7.692 & 3,1,1 & 7.669 &3,2,2 &7.687 &3,3,3 &7.709\\
& 3,2,0 &7.726 & 3,2,1 & 7.716 &3,3,2 &7.743 &3,4,3 &7.773\\
& 3,3,0 &7.768 & 3,3,1 & 7.760 &3,4,2 &7.796 &3,5,3 &7.833\\
& 3,4,0 &7.816 & 3,4,1 & 7.810 &3,5,2 &7.852 &3,6,3 &7.896\\
& 3,5,0 &7.870 & 3,5,1 & 7.864 &3,6,2 &7.912 &3,7,3 &7.961\\
\hline
\end{tabular}

\newpage

\end{document}